\documentclass[a4paper]{jpconf}
\usepackage{graphicx}
\usepackage[square,sort&compress]{natbib}

\begin{document}
\title{Analysis of the Hydrogen-rich Magnetic White Dwarfs in the SDSS}

\author{Baybars K\"ulebi$^{1}$, Stefan Jordan$^{1}$, Fabian Euchner$^{2}$,\\
	Heiko Hirsch$^{3}$, and Wolfgang L\"offler$^{1}$}
\address{$^1$Astronomisches Rechen-Institut, M\"onchhofstrasse 12-14,
	69120 Heidelberg,
	Germany }
\address{$^2$Swiss Seismological Service, ETH Zurich,
	Sonneggstrasse 5, 8092 Zurich, Switzerland}
\address{$^3$Dr.-Remeis-Sternwarte, Bamberg, Sternwartstrasse 7, D-96049 Bamberg}

\ead{bkulebi@ari.uni-heidelberg.de}

\begin{abstract}
We have calculated  optical spectra of hydrogen-rich (DA) white dwarfs with magnetic field strengths between 1 MG and
1000 MG for temperatures between 7000 K and 50000 K. Through a least-squares minimization scheme with an
evolutionary algorithm, we have analyzed the spectra of 114 magnetic DAs from the SDSS (95 previously
published plus 14 newly discovered within SDSS, and five discovered by SEGUE). Since we were limited to a single spectrum for each object we used only centered magnetic dipoles or dipoles which were shifted along the magnetic dipole axis. We also statistically  investigated the distribution of magnetic-field strengths and geometries of our sample.

\end{abstract}

\section{Introduction}
\label{Introduction}

The Sloan Digital Sky Survey (SDSS) project covering $10^4$ ${\rm deg}^2$ of the sky has discovered thousands of new white dwarfs, among them 105 with magnetic fields
(MWDs) \citep{Gaeansickeetal02, Schmidtetal03, Vanlandinghametal05}.White dwarfs with magnetic fields between $10^4$ and $ 10^9$ G are thought to represent more than 10\% of the total population of white dwarfs \citep{Liebertetal03}. The SDSS Data Release 3 (DR3, $\approx 4200\, {\rm deg}^2$) increased the number of known magnetic white dwarfs from 65 \citep{WickramasingheFerrario00,Jordan01} to 170 \citep{Kawkaetal07}. 

In this work we present the re-analysis of the 95 DA MWDs discovered by \citet{Schmidtetal03} and \citet{Vanlandinghametal05},  plus the analysis of 19 additional objects from SDSS up to DR6\footnote{http://www.sdss.org/dr6/} ($9583\,{\rm deg}^2$). 

\citet{Schmidtetal03} and \citet{Vanlandinghametal05} determined the field strengths  and the  inclinations of magnetic dipoles by comparing visually the observed spectra  with model spectra calculated
using a simplified radiation transfer code \citep{LatterSchmidt87}.
Their analyses resulted in dipolar field strengths for the  SDSS MWDs  between $1.5$\,MG and $\sim1000$\,MG. From their sample of 105 MWDs, 97 were classified as DAs.

\section{SDSS data}
\label{SDSS data}
 The spectroscopic targets of the SDSS are  selected based on color criteria  optimized for galaxies and quasars. However, follow-up spectroscopy of many stars is also performed with the twin dual beam spectrographs (3900\,-\,6200 and 5800\,-\,9200 \AA, $\lambda /d\lambda \sim 1800$), in particular of blue objects like white dwarfs and hot subdwarfs \citep{Harrisetal03, Kleinmanetal04}. Since the energy distribution  of strongly magnetic white dwarfs can differ from nonmagnetic ones, MWDs are not only found in the SDSS color categories for white dwarfs or blue horizontal-branch stars, but may also fall into the color categories for quasars (QSOs), ``serendipitous blue objects'', and hot subdwarfs. Based on their colors, objects are assigned to fibers for spectroscopic investigation (for spectroscopic target selection, see \cite{Stoughton02}). 

In order to identify magnetic white dwarfs among the white dwarfs or other categories, different techniques were used: From the sample of white dwarfs, selected by color cuts, \citet{Gaeansickeetal02} and \citet{Schmidtetal03} used visual inspection. \citet{Vanlandinghametal05} inspected only those objects visually for which bad $\chi^2$ were obtained by their \textit{autofit} process. This procedure selects in particular white dwarfs with  magnetic fields above 3\,MG and misses objects with weaker magnetic fields.

In addition to these already known objects, we have analyzed data of nineteen additional SDSS objects discovered to be magnetic due to suspicious radial velocity measurements by the HYPERMUCHFUSS project (see, Geier et al, these proceedings and Tillich et al, these proceedings). The one-dimensional spectra which we used in this work  were generated by SDSS's spectroscopic pipeline \texttt{spectro2d} and downloaded from Data Archive Server.

\section{Analysis}
\label{Analysis}

\begin{figure}
   \centering
   \includegraphics[width=0.95\textwidth]{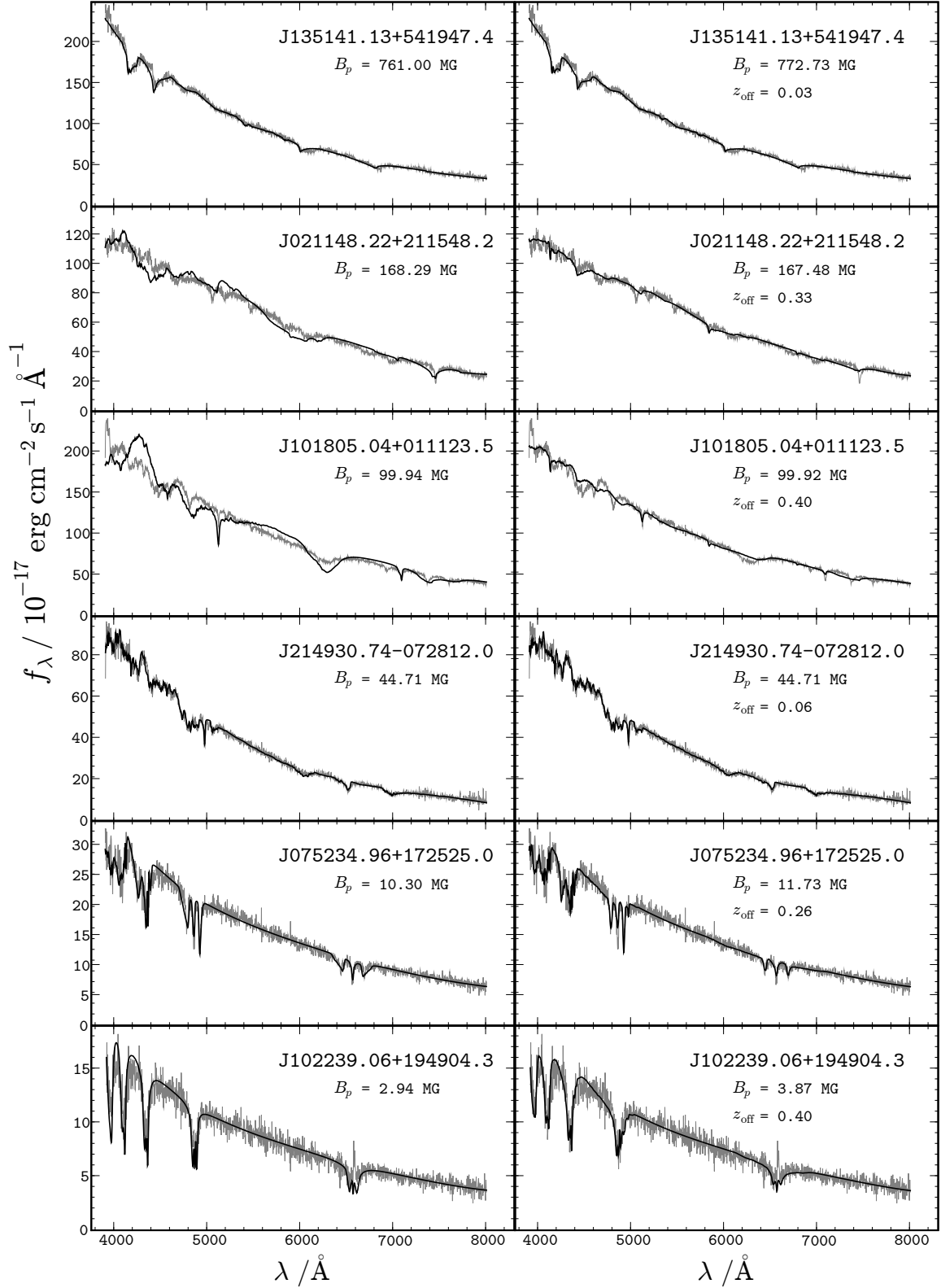}
  \caption{Representative sample of fits of observed spectra of MWDs from the SDSS to centered magnetic dipoles with 
   a polar field strength  $B_p$ (left) and dipoles shifted by $z_{\rm off}$ stellar radii along the dipole axis
  (right).}
  \label{fig:bestfit}
\end{figure}

  In order to increase efficiency, we pre-computed a three-dimensional grid of Stokes $I$ and $V$ model spectra with  effective  temperature $7000\,{\rm K}\le T_{\rm eff} \le 50000\,{\rm K}$, magnetic field strength 
$1\,{\rm MG}\le B\le 1\,{\rm GG}$, and 18  different  directions $\psi$ relative to the line of sight as the independent variables (9 entries, equally spaced in   $\cos \psi$)
using the radiative transfer code for magnetized white dwarf atmospheres  \cite[see][]{Jordan92,JordanSchmidt03}.
 All spectra are calculated for a surface gravity of \mbox{$\log g = 8$}.
Since no polarization information is available for the SDSS, our analysis is limited to the flux spectra (Stokes parameter $I$). 

The magnetic field geometry of the magnetic white dwarfs was determined with a modified version of the code developed by \citet{Euchneretal02}. 
This code calculates the total flux (and circular polarization) spectra for an arbitrary magnetic field topology by adding up appropriately weighted model spectra for a large number of surface elements. Complex magnetic field geometries are accounted for by a multipole expansion of the scalar magnetic potential, but in our paper only centered and offset dipoles were considered.
The observed spectra are  fitted using an evolutionary algorithm  \citep{Rechenberg94} with a least-squares quality function.

Due to the lack of a consistent theory for Zeeman and Stark broadening, the latter is only taken into account by a rather simple approximation \citep{Jordan92} so that systematic uncertainties, particularly in the
low-field regime ($\le 5$\,MG), are unavoidable.  Consequently, effective temperatures and surface gravities derived from fitting the Balmer lines alone are less reliable than in the case of non-magnetic white dwarfs. This may also result in disagreements with temperatures estimates derived from the continuum slope.

Our fitting procedure had to adjust three  or four  free parameters: the magnetic dipole field strength $B_d$, the effective temperature $T_{\rm eff}$, the inclination of the dipole axis $i$, %\citep[see][]{Bergeronetal92,Finleyetal97}
and an offset along the magnetic axis $z_{\rm off}$, if offset dipoles are used.  $T_{\rm eff}$ needed to be independently calculated for our purposes. For the 95 MWDs  analyzed, we used the literature values for $T_{\rm eff}$ which were determined by comparison to the theoretical non-magnetic DA colors in the $u-g$ vs $g-r$ plane \citep{Schmidtetal03,Vanlandinghametal05}. The temperature of the new MWDs presented in Table\,\ref{tab:newMWD} were estimated by the synthetic SDSS color-color diagrams by \citet{HolbergBergeron06}.

All fits resulted in  reduced  $\chi^2$ values between 0.8 and  3.0 except for some high-field objects.  In Fig.\,\ref{fig:bestfit} we show fits of 6 MWD spectra as an example.

\begin{figure}[t]
   \centering
   \includegraphics[width=0.56\textwidth]{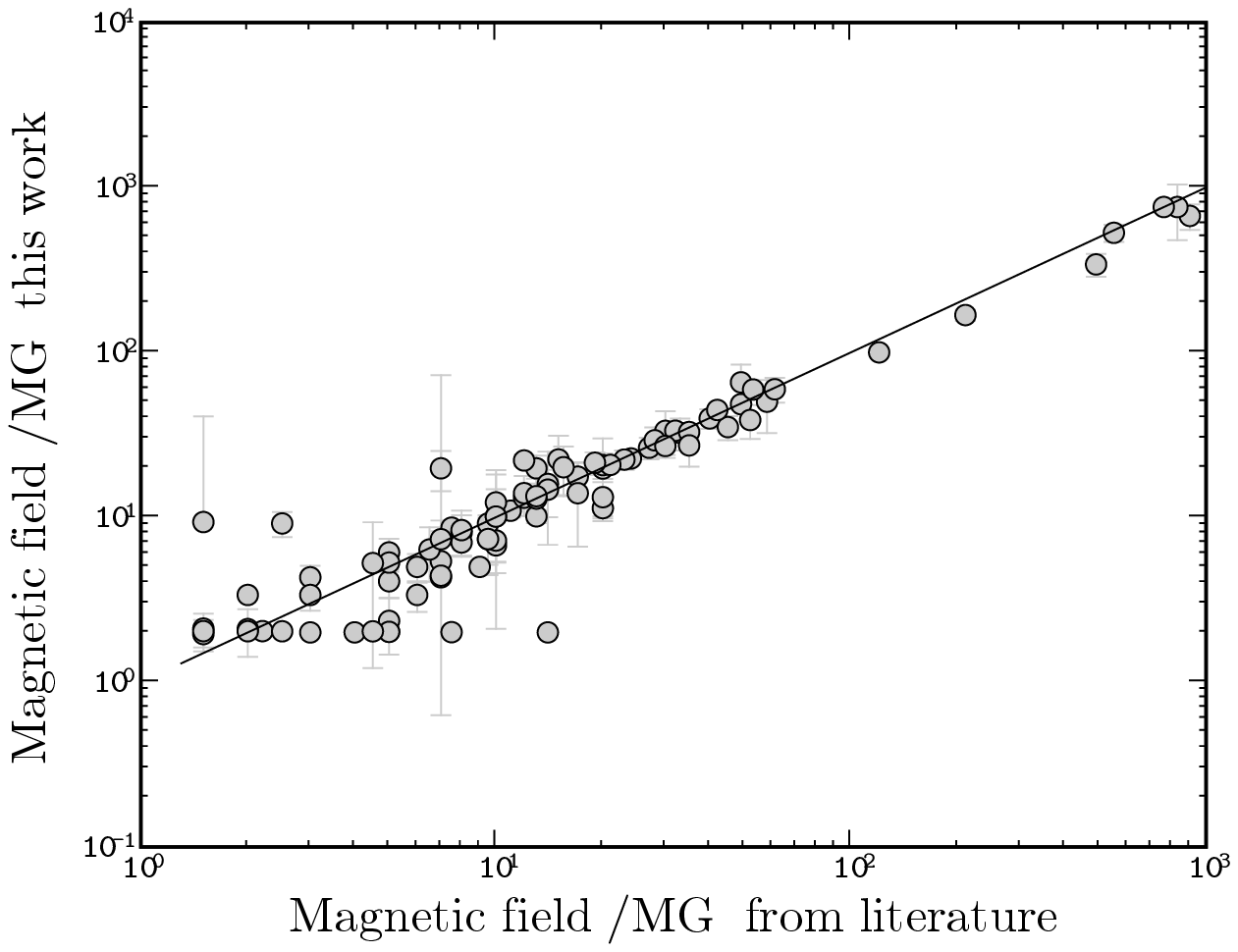}
\hfill
\begin{minipage}[b]{0.4\textwidth}
  \caption{Comparison of dipole magnetic field fit values in this work versus \citet{Schmidtetal03}, \citet{Vanlandinghametal05}.}
\end{minipage}
  \label{fig:comparison}
\end{figure}

%\end{document}
\begin{table}
\begin{tabular}{l|rc|rcc|r}
\hline
\hline \noalign{\smallskip}
MWD (SDSS+) & $B_p$ / MG  &  $i$ / deg& $B_{\rm off}$ / MG  & $z_{\rm off}$ / $r_{\rm WD}$ & $i$ / deg & $T_{\rm{eff}}$\\ \noalign{\smallskip} \hline \noalign{\smallskip}
J031824.19+422651.0	&	$10.12	\pm	$0.10	&$54.6\pm4.7$	&	$10.77	\pm	$0.10	&	$0.29	\pm	$0.05	&$61.1\pm10.0$	&	10500	\\
J075234.96+172525.0	&	$10.30	\pm	$1.23	&$72.4\pm22.4$	&	$11.73	\pm	$1.05	&	$0.26	\pm	$0.05	&$58.7\pm38.1$	&	9000	\\
J083945.56+200015.7	&	$3.38	\pm	$0.49	&$48.6\pm7.7$	&	$2.15	\pm	$0.10	&	$0.29	\pm	$0.08	&$49.9\pm90\footnotemark[1]$	&	15000	\\
J085106.12+120157.8	&	$2.03	\pm	$0.10	&$81.9\pm90\footnotemark[1]$	&	$2.47	\pm	$0.10	&	$0.35	\pm	$0.06	&$72.8\pm18.8$	&	11000	\\
J085523.87+164059.0	&	$12.23	\pm	$2.92	&$48.6\pm8.6$	&	$7.86	\pm	$1.63	&	$0.36	\pm	$0.06	&$10.8\pm6.1$	&	15500	\\
J091833.32+205536.9	&	$2.04	\pm	$0.10	&$87.2\pm41.9$	&	$2.66	\pm	$1.71	&	$0.39	\pm	$0.17	&$70.3\pm61.9$	&	14000	\\
J100657.51+303338.1	&	$1.00	\pm	$0.10	&$82.5\pm30.8$	&	$1.30	\pm	$1.23	&	$0.37	\pm	$0.39	&$14.3\pm13.1$	&	10000	\\
J102239.06+194904.3	&	$2.94	\pm	$0.71	&$49.0\pm13.0$	&	$3.87	\pm	$1.11	&	$0.40	\pm	$0.07	&$51.3\pm40.3$	&	9000	\\
J112257.10+322327.8	&	$11.38	\pm	$3.42	&$49.0\pm12.3$	&	$7.46	\pm	$1.68	&	$0.37	\pm	$0.11	&$4.2\pm6.3$	&	12500	\\
J125434.65+371000.1	&	$4.10	\pm	$0.35	&$41.9\pm19.2$	&	$4.89	\pm	$0.42	&	$0.40	\pm	$0.03	&$50.1\pm21.0$	&	10000	\\
J125715.54+341439.3	&	$11.45	\pm	$0.71	&$0.5\pm0.6$	&	$13.70	\pm	$1.69	&	$0.07	\pm	$0.02	&$7.7\pm12.0$	&	8500	\\
J134820.79+381017.2	&	$13.65	\pm	$2.66	&$89.4\pm90\footnotemark[1]$	&	$14.45	\pm	$4.65	&	$0.22	\pm	$0.04	&$54.8\pm25.3$	&	35000	\\
J140716.66+495613.7	&	$12.49	\pm	$6.20	&$88.1\pm90\footnotemark[1]$	&	$13.20	\pm	$4.21	&	$0.24	\pm	$0.10	&$63.3\pm81.1$	&	20000	\\
J141906.19+254356.5	&	$2.03	\pm	$0.10	&$81.2\pm8.7$	&	$2.56	\pm	$0.10	&	$0.38	\pm	$0.03	&$54.8\pm10.4$	&	9000	\\
J143019.05+281100.8	&	$9.34	\pm	$1.44	&$5.6\pm4.5$	&	$6.25	\pm	$0.75	&	$0.16	\pm	$0.03	&$5.6\pm4.5$	&	9000	\\
J151130.17+422023.0	&	$22.40	\pm	$9.41	&$48.6\pm19.5$	&	$8.37	\pm	$1.07	&	$0.31	\pm	$0.06	&$5.8\pm21.1$	&	9500	\\
J202501.10+131025.6	&	$10.10	\pm	$1.76	&$68.5\pm9.1$	&	$10.72	\pm	$1.71	&	$0.29	\pm	$0.04	&$53.7\pm9.0$	&	17000	\\
J220435.05+001242.9	&	$1.02	\pm	$0.10	&$71.2\pm90\footnotemark[1]$	&	$2.50	\pm	$5.47	&	$0.36	\pm	$0.69	&$3.1\pm13.6$	&	22000	\\
J225726.05+075541.7	&	$16.17	\pm	$2.81	&$74.9\pm16.1$	&	$17.39	\pm	$3.21	&	$0.15	\pm	$0.05	&$78.5\pm34.8$	&	40000	\\

\noalign{\smallskip} \hline
\end{tabular}
\caption{Best fit parameter values for centered dipole and offset dipole models for the objects discovered by HYPERMUCHFUSS. \\\footnotemark[1]Inclination errors are estimated as large, for explanation see Sec. 4}\label{fitvalues}
\label{tab:newMWD}
\end{table}

\begin{figure}[h]
\begin{minipage}{16pc}
 \includegraphics[width=\textwidth]{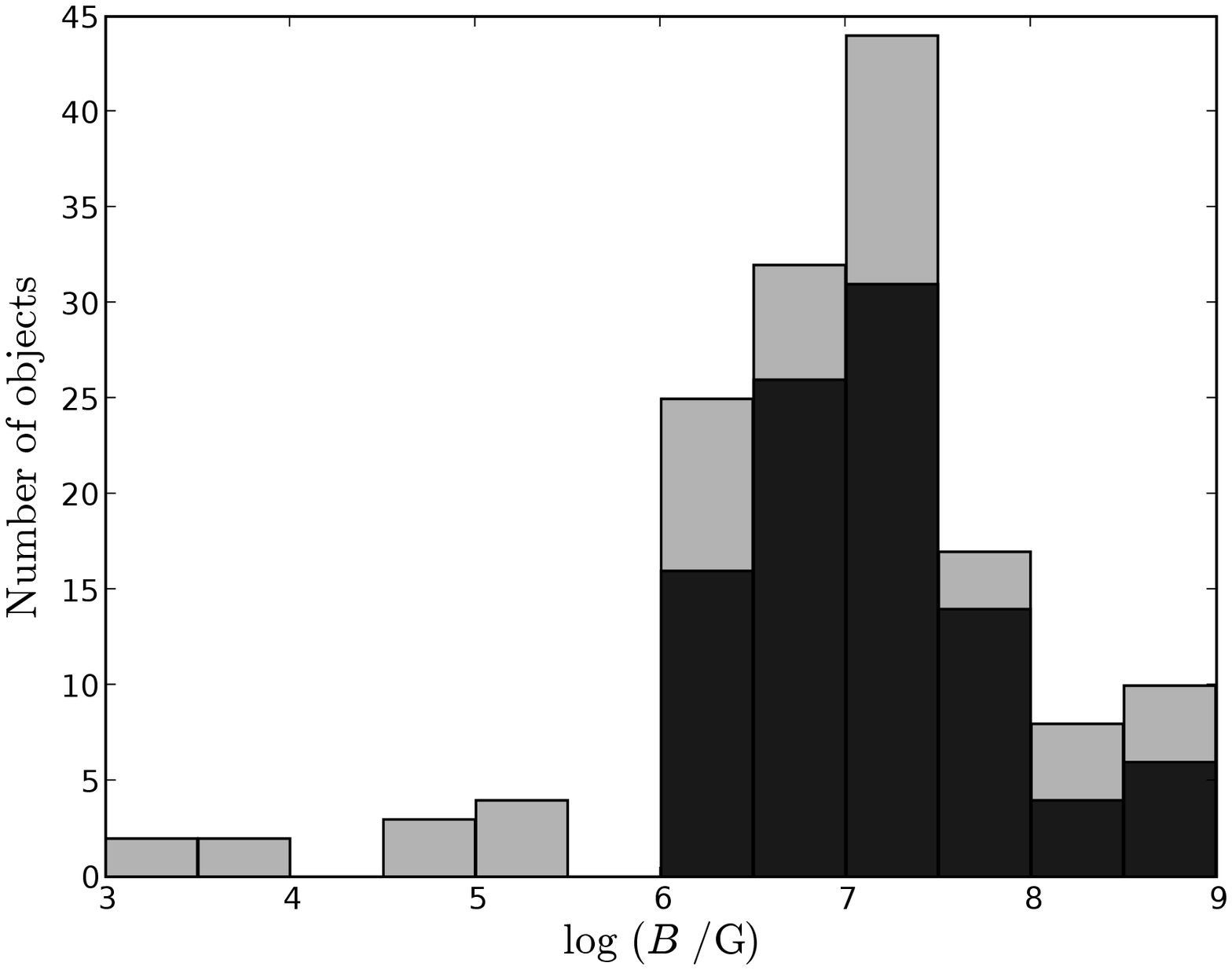}
  \caption{\label{fig:histogram}Histogram of magnetic white dwarfs in equal intervals of $\log B$. Gray columns represent the number of all DA MWDs and black shades represent the the contribution of SDSS to DA MWDs.}\end{minipage}\hspace{2pc}%
\begin{minipage}{20pc}
 \includegraphics[width=0.62\textwidth, bb = 75 180 450 490]{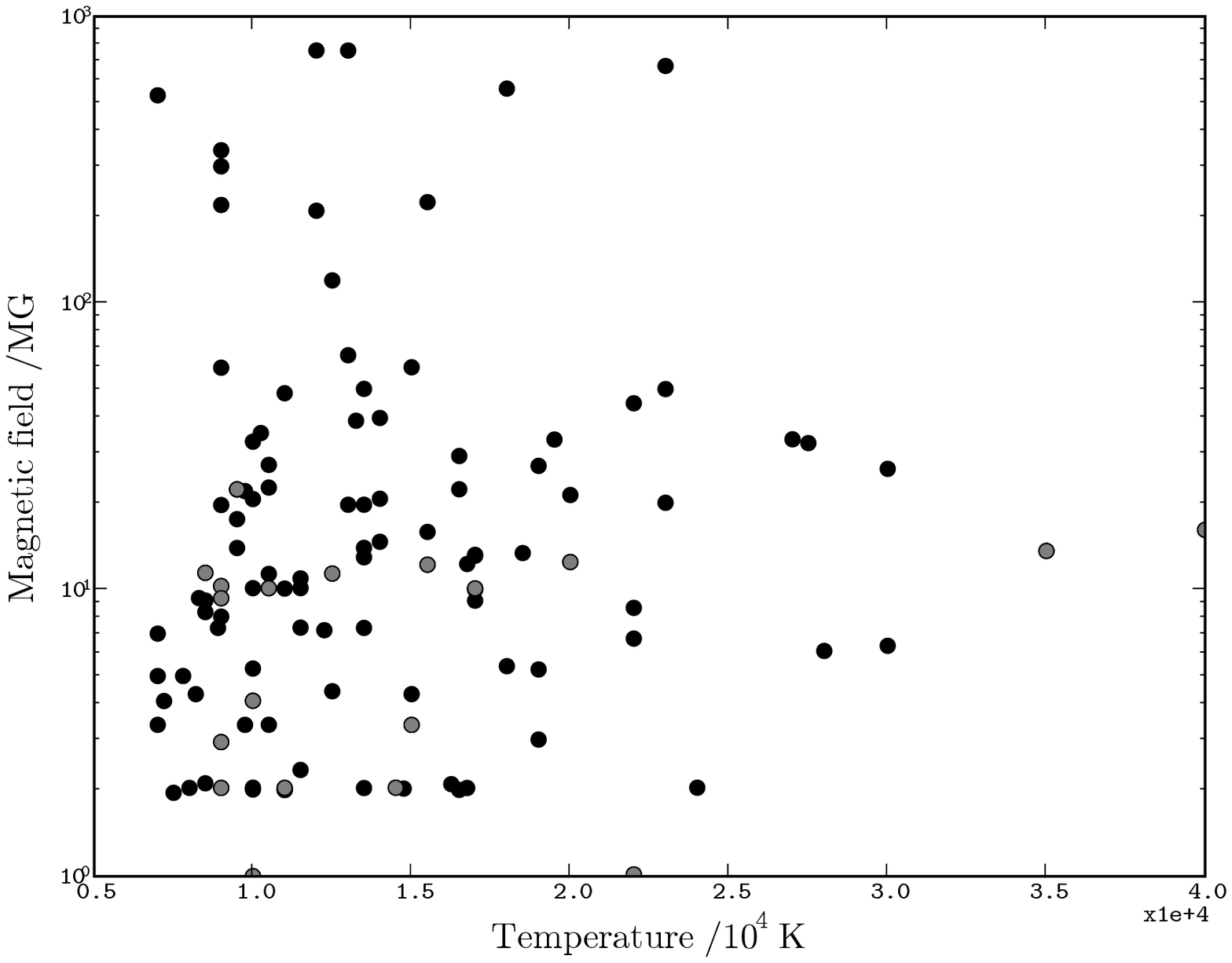}
  \caption{\label{fig:bvst}Scatter plot of dipole magnetic field value vs. temperature in this work. The random distribution of field strengths with respect to the age indicator temperature is consistent with a long decay timescale of MWDs. Gray dots indicate the objects from HYPERMUCHFUSS.}
\end{minipage} 

\end{figure}

%\section{Results}
%\label{Results}
\section{Discussion}
\label{Individual}
Three objects analyzed by \citet{Schmidtetal03} and \citet{Vanlandinghametal05} are omitted in this work. \textit{SDSSJ05959.56+433521.3, G111-49} was listed by \citet{Schmidtetal03} as a magnetic DA, but is a Carbon-rich (DC) MWD \citep{Putney95}; 
\textit{SDSSJ084716.21+148420.4} is a DAH+DB binary, hence we were unable to analyze these two objects with our code. Finally, neither the coordinates nor fiber identifier of \textit{SDSSJ234605.44+385337.7} mentioned by \citet{Vanlandinghametal05} was accessible through the  SDSS database.

Emission lines were found in \textit{SDSSJ102220.69+272539.8} and \textit{SDSSJ\-102239.06\-+194904.3}  (the latter is shown in  Fig.\,\ref{fig:bestfit}), very similar to \textit{SDSSJ121209.31\-+013627.7} which could indicate that these objects may be EF Eri like, magnetic cataclysmic variables with a brown dwarf companion \citep{Burleighetal07}.

The spectra of the high-field objects \textit{SDSSJ224741.41\-+145638.8} and \textit{SDSSJ101805.04\-+01123.5} (PG\,1015+014, shown in  Fig.\,\ref{fig:bestfit}) do not fit particularly well. At higher field strengths ($>\,50$\,MG) the spectra
become very sensitive to the details of the magnetic field geometry, as was demonstrated by \citet{Euchneretal02,Euchneretal05,Euchneretal06}.  The deviations of the observed spectra from our theoretical spectra assuming  (offset) dipole models hint therefore to a magnetic field geometry that is more complex than a shifted dipole. A more comprehensive analysis of PG\,1015+014 \citep{Euchneretal06} showed that individually tilted and off-centered zonal multipole components with field strengths between 50-90 MG are needed to represent the global magnetic field, consistent with our analysis. On the other side, some high-field objects in our sample, like  \textit{SDSSJ135141.13\-+541947} (Fig.\,\ref{fig:bestfit}) are well fitted.

SDSS has proven to be extremely important in increasing the total number of MWDs. Out of 8000 WDs (DR4; \cite{Eisensteinetal06}), a total 139 MWDs were discovered \cite[DR6;][]{Kulebietal}, still a very low percentage compared to the sample of MWDs in the solar neighborhood (13$\% \pm4\%$; \cite{Kawkaetal07}). Note, however, that no systematic search for MWDs in DR4-DR6 was performed yet.
SDSS is designed to investigate galaxies and quasars, thus stars are not its primary targets. Formerly the WD acquisition efficiency of the SDSS was partly defined by the probability of WD spectra to be target labelled as QSO. With each new data release, target selection criteria has improved in separating the QSO from non-QSO targets. Unfortunately the improvement in extragalatic target selection impairs SDSSs capability to detect serendipitous MWDs. This was examplified by the inability to recover most of the formerly known MWDs in regions of the sky covered in DR2-3 by \citet{Vanlandinghametal05}.

Overall, SDSS has nearly tripled the number of MWDs, conversely the completeness of the total MWD population is affected significantly by the SDSS biases because of this high impact of SDSS. The most important selection effect 
arises from the priority selection of the spectroscopic targeting, which is determined from color categories of each object \citep{Stoughton02}. Due to unconventional spectrum and absorption features, MWDs with high field strengths have a tendency to be labeled as interesting objects (e.g., BLUE\_SERENDIPITY, which is an object in undocumented coordinates with blue colors). This contributes as a positive bias for MWDs with high magnetic field strengths ($B >$ 100\,MG). On the other side, seven objects from the  HYPERMUCHFUSS contributed to the  $B\le 3.5$\,MG  range .
This limit was noted to be a negative bias  by \cite{Vanlandinghametal05}, due to the insensitivity of the \textit{autofit} process. Since spectra from objects with such  low fields only slightly differ from the non-magnetic cases, they were not found by the automatic fitting process and therefore no visual inspection was triggered. Since some HYPERMUCHFUSS  objects were found in DR1 to DR3, the radial velocity method has proven to be complementary in finding some of the missing low-field MWDs. 

These selection effects do not apply to the second phase of the SDSS, Sloan Extension for Galactic Understanding and Exploration (SEGUE), which is targeting the stellar population specifically. This includes the white dwarfs and blue horizontal-branch stars\footnote{\rm{http://segue.uchicago.edu/targetsel.html\#v4.2}} (that are also in MWD color criteria). In our sample, five out of the nineteen new MWDs are from SEGUE.

\section{Conclusion}
Overall, our results are consistent with the former analyses of MWDs (see Fig.\,\ref{fig:comparison}), which shows that simple atmosphere models with preassumed dipole magnetic values are good approximations for these objects, if only single-phase spectroscopy without polarization information is considered. Our method is able to account for a diversity in low magnetic field strengths ($B\,<\,20$\,MG). This is apparent in the deviation of our best-fit magnetic field strength values from literature values in the horizontal $\sim$2\,MG axis of Figure\,\ref{fig:comparison}. In many cases offset dipole models resulted in significantly better fits than the models with centered dipoles, however, at smaller field strength this can be in part due to the simplifications in the treatment of Stark broadening.

The distribution of the magnetic field strengths of the MWDs from the \citet{Schmidtetal03} sample had  a maximum around $\sim$5\,-\,30 MG. We have updated the values of DA white dwarfs and created a histogram of all known magnetic DA WDs  (Fig.\,\ref{fig:histogram}) and added results for the new objects. The magnetic field strengths of non-SDSS MWDs are from the compilation by \citet{Kawkaetal07}. In spite of the SDSS biases, the same $\sim$5\,-\,30 MG peak as in \citet{Schmidtetal03} is apparent in Fig.\,\ref{fig:histogram}.

The dipole magnetic field ohmic decay timescale is $10^{10}$ yr. Even the higher multipoles can live for such a long period of time (see, e.g. \cite{Muslimovetal95}). Therefore, no significant correlation between temperature and magnetic field strength is expected if temperature is assumed as an indicator of age (Fig.\,\ref{fig:bvst}). This lack of correlation supports the fossil ancestry of these fields inherited from earlier stages of stellar evolution. 

High field MWDs are thought to be remnants of magnetic Ap and Bp stars. If flux conservation is assumed, the distribution of the polar field strengths of high field MWDs should be largest in the interval \mbox{50--500\,MG}. In our sample, objects with magnetic field strengths lower than 50 MG are more numerous than the objects with higher magnetic field strengths (see Fig.\,\ref{fig:histogram}), part of this effect is due to our biases (see Section \ref{SDSS data}).  Nevertheless it is consistent with previous results and supports the hypothesis that magnetic fossil fields from Ap/Bp stars only are not sufficient to produce high field MWDs \citep{WickramasingheFerrario05MN}. \citet{Auriereetal07} argued that dipole magnetic field strengths of magnetic Ap/Bp stars have a ``magnetic threshold'' due to large scale stability conditions, and this results in a steep decrease in the number of magnetic Ap/Bp stars below polar magnetic fields of 300\,G. 

A possible progenitor population for MWDs with dipolar field strengths below 50 MG is the currently unobserved population of A and B stars with magnetic field strengths of 10\,-\,100\,G. \citet{WickramasingheFerrario05MN} suggested that if $\sim40\%$ of A/B stars have magnetism, this would be sufficient to explain the observed distribution of MWDs. However, the existence of this population seems to be highly unlikely since investigations of \citet{Shorlinetal02} and \citet{Bagnuloetal06} for magnetism in this population yielded null results, for median errors of 15\,-\,50\,G and 80\,G respectively. Another candidate group is the yet undetected magnetic F stars for these MWDs with lower field strengths \citep{Schmidtetal03}. But this conclusion is strongly affected by SDSS MWD discovery biases.

\ack{
      Part of this work is supported by the DLR under grant 50 OR 0802. Funding for the creation and distribution of the SDSS archive has been provided by the Alfred P. Sloan Foundation, the SDSS member institutions, the National Aeronautics and Space Administration, the National Science Foundation, the U.S. Department of Energy, Monbusho, and the Max Planck Society. The SDSS World Wide Web site is http://www.sdss.org/.}

%\section*{References}
%\bibliographystyle{aa}
%\bibliography{bibdata}
\bibliography{kulebi.bbl}

\providecommand{\newblock}{}
\begin{thebibliography}{10}
\expandafter\ifx\csname url\endcsname\relax
  \def\url#1{{\tt #1}}\fi
\expandafter\ifx\csname urlprefix\endcsname\relax\def\urlprefix{URL }\fi
\providecommand{\eprint}[2][]{\url{#2}}

\bibitem[{{G{\"a}nsicke} {et~al.}(2002){G{\"a}nsicke}, {Euchner}, \&
  {Jordan}}]{Gaeansickeetal02}
{G{\"a}nsicke} B~T, {Euchner} F and {Jordan} S 2002 {\em \aap\/} {\bf 394}
  957--963 (\textit{Preprint} \eprint{arXiv:astro-ph/0208454})

\bibitem[{{Schmidt} {et~al.}(2003){Schmidt}, {Harris}, {Liebert}, {Eisenstein},
  {Anderson}, {Brinkmann}, {Hall}, {Harvanek}, {Hawley}, {Kleinman}, {Knapp},
  {Krzesinski}, {Lamb}, {Long}, {Munn}, {Neilsen}, {Newman}, {Nitta},
  {Schlegel}, {Schneider}, {Silvestri}, {Smith}, {Snedden}, {Szkody}, \&
  {Vanden Berk}}]{Schmidtetal03}
{Schmidt} G~D, {\em et~al.~} 2003 {\em \apj\/}
  {\bf 595} 1101--1113 (\textit{Preprint} \eprint{arXiv:astro-ph/0307121})

\bibitem[{{Vanlandingham} {et~al.}(2005){Vanlandingham}, {Schmidt},
  {Eisenstein}, {Harris}, {Anderson}, {Hall}, {Liebert}, {Schneider},
  {Silvestri}, {Stinson}, \& {Wolfe}}]{Vanlandinghametal05}
 {Vanlandingham} K, {\em et~al.~} 
  2005 {\em \aj\/} {\bf 130} 734--741 (\textit{Preprint}
  \eprint{arXiv:astro-ph/0505085})

\bibitem[{{Liebert} {et~al.}(2003){Liebert}, {Bergeron}, \&
  {Holberg}}]{Liebertetal03}
{Liebert} J, {Bergeron} P and {Holberg} J~B 2003 {\em \aj\/} {\bf 125} 348--353
  (\textit{Preprint} \eprint{arXiv:astro-ph/0210319})

\bibitem[{{Wickramasinghe} \& {Ferrario}(2000)}]{WickramasingheFerrario00}
{Wickramasinghe} D~T and {Ferrario} L 2000 {\em \pasp\/} {\bf 112} 873--924

\bibitem[{{Jordan}(2001)}]{Jordan01}
{Jordan} S 2001 {\em 12th European Workshop on White Dwarfs\/} ({\em
  Astronomical Society of the Pacific Conference Series\/} vol 226) ed
  {Provencal} J~L, {Shipman} H~L, {MacDonald} J and {Goodchild} S pp 269--+

\bibitem[{{Kawka} {et~al.}(2007){Kawka}, {Vennes}, {Schmidt}, {Wickramasinghe},
  \& {Koch}}]{Kawkaetal07}
{Kawka} A, {Vennes} S, {Schmidt} G~D, {Wickramasinghe} D~T and {Koch} R 2007
  {\em \apj\/} {\bf 654} 499--520 (\textit{Preprint}
  \eprint{arXiv:astro-ph/0609273})

\bibitem[{{Latter} {et~al.}(1987){Latter}, {Schmidt}, \&
  {Green}}]{LatterSchmidt87}
{Latter} W~B, {Schmidt} G~D and {Green} R~F 1987 {\em \apj\/} {\bf 320}
  308--314

\bibitem[{{Harris} {et~al.}(2003){Harris}, {Liebert}, {Kleinman}, {Nitta},
  {Anderson}, {Knapp}, {Krzesi{\'n}ski}, {Schmidt}, {Strauss}, {Vanden Berk},
  {Eisenstein}, {Hawley}, {Margon}, {Munn}, {Silvestri}, {Smith}, {Szkody},
  {Collinge}, {Dahn}, {Fan}, {Hall}, {Schneider}, {Brinkmann}, {Burles},
  {Gunn}, {Hennessy}, {Hindsley}, {Ivezi{\'c}}, {Kent}, {Lamb}, {Lupton},
  {Nichol}, {Pier}, {Schlegel}, {SubbaRao}, {Uomoto}, {Yanny}, \&
  {York}}]{Harrisetal03}
{Harris} H~C, {\em et~al.~} 2003 {\em
  \aj\/} {\bf 126} 1023--1040 (\textit{Preprint}
  \eprint{arXiv:astro-ph/0305347})

\bibitem[{{Kleinman} {et~al.}(2004){Kleinman}, {Harris}, {Eisenstein},
  {Liebert}, {Nitta}, {Krzesi{\'n}ski}, {Munn}, {Dahn}, {Hawley}, {Pier},
  {Schmidt}, {Silvestri}, {Smith}, {Szkody}, {Strauss}, {Knapp}, {Collinge},
  {Mukadam}, {Koester}, {Uomoto}, {Schlegel}, {Anderson}, {Brinkmann}, {Lamb},
  {Schneider}, \& {York}}]{Kleinmanetal04}
{Kleinman} S~J, {\em et~al.~}
  2004 {\em \apj\/} {\bf 607} 426--444 (\textit{Preprint}
  \eprint{arXiv:astro-ph/0402209})

\bibitem[{{Stoughton} {et~al.}(2002)}]{Stoughton02}
{Stoughton} C, {\em et~al.~} 2002 {\em \aj\/} {\bf 123} 485--548

\bibitem{Tillich08}
{Tillich} A, {Heber} U and {Hirsch} H~A 2008 {\em Hot Subdwarf Stars and
  Related Objects\/} ({\em Astronomical Society of the Pacific Conference
  Series\/} vol 392) ed {Heber} U, {Jeffery} C~S and {Napiwotzki} R pp 175--+

\bibitem{Jordan92}
{Jordan} S 1992 {\em \aap\/} {\bf 265} 570--576

\bibitem{JordanSchmidt03}
{Jordan} S and {Schmidt} H 2003 {\em Stellar Atmosphere Modeling\/} ({\em
  Astronomical Society of the Pacific Conference Series\/} vol 288) ed {Hubeny}
  I, {Mihalas} D and {Werner} K pp 625--+

\bibitem[{{Euchner} {et~al.}(2002){Euchner}, {Jordan}, {Beuermann},
  {G{\"a}nsicke}, \& {Hessman}}]{Euchneretal02}
{Euchner} F, {Jordan} S, {Beuermann} K, {G{\"a}nsicke} B~T and {Hessman} F~V
  2002 {\em \aap\/} {\bf 390} 633--647 (\textit{Preprint}
  \eprint{arXiv:astro-ph/0205294})

\bibitem[{{Rechenberg}(1994)}]{Rechenberg94}
{Rechenberg} I 1994 {\em Werkstatt Bionik und Evolutionstechnik No.~1
  (Stuttgart: frommann-holzboog\/}

\bibitem[{{Holberg} \& {Bergeron}(2006)}]{HolbergBergeron06}
{Holberg} J~B and {Bergeron} P 2006 {\em \aj\/} {\bf 132} 1221--1233

\bibitem[{{Putney}(1995)}]{Putney95}
{Putney} A 1995 {\em \apjl\/} {\bf 451} L67+

\bibitem[{{Burleigh} {et~al.}(2007){Burleigh}, {Marsh}, {G{\"a}nsicke}, {Goad},
  {Dhillon}, {Littlefair}, {Wells}, {Dobbie}, {Farihi}, {Bannister},
  {Casewell}, {Hurkett}, {Martindale}, {Roche}, \& {Lewis}}]{Burleighetal07}
{Burleigh} M~R, {Marsh} T~R, {G{\"a}nsicke} B~T, {Goad} M~R, {Dhillon} V~S,
  {Littlefair} S~P, {Wells} M, {Dobbie} P~D, {Farihi} J, {Bannister} N~P,
  {Casewell} S~L, {Hurkett} C~P, {Martindale} A, {Roche} P and {Lewis} F 2007
  {\em 15th European Workshop on White Dwarfs\/} ({\em Astronomical Society of
  the Pacific Conference Series\/} vol 372) ed {Napiwotzki} R and {Burleigh}
  M~R pp 477--+

\bibitem[{{Euchner} {et~al.}(2005){Euchner}, {Reinsch}, {Jordan}, {Beuermann},
  \& {G{\"a}nsicke}}]{Euchneretal05}
{Euchner} F, {Reinsch} K, {Jordan} S, {Beuermann} K and {G{\"a}nsicke} B~T 2005
  {\em \aap\/} {\bf 442} 651--660 (\textit{Preprint}
  \eprint{arXiv:astro-ph/0507631})

\bibitem[{{Euchner} {et~al.}(2006){Euchner}, {Jordan}, {Beuermann}, {Reinsch},
  \& {G{\"a}nsicke}}]{Euchneretal06}
{Euchner} F, {Jordan} S, {Beuermann} K, {Reinsch} K and {G{\"a}nsicke} B~T 2006
  {\em \aap\/} {\bf 451} 671--681 (\textit{Preprint}
  \eprint{arXiv:astro-ph/0602112})

\bibitem[{{Eisenstein} {et~al.}(2006){Eisenstein}, {Liebert}, {Harris},
  {Kleinman}, {Nitta}, {Silvestri}, {Anderson}, {Barentine}, {Brewington},
  {Brinkmann}, {Harvanek}, {Krzesi{\'n}ski}, {Neilsen}, {Long}, {Schneider}, \&
  {Snedden}}]{Eisensteinetal06}
{Eisenstein} D~J, {\em et~al.~} 2006 {\em \apjs\/} {\bf 167} 40--58 (\textit{Preprint}
  \eprint{arXiv:astro-ph/0606700})

\bibitem[{{K\"ulebi} {et~al.}(in preparation){K\"ulebi}, {Jordan}, {Euchner},
  \& {Hirsch}}]{Kulebietal}
{K\"ulebi} B, {Jordan} S, {Euchner} F, {Hirsch} H and {G\"ansicke} B in preparation

\bibitem[{{Muslimov} {et~al.}(1995){Muslimov}, {van Horn}, \&
  {Wood}}]{Muslimovetal95}
{Muslimov} A~G, {van Horn} H~M and {Wood} M~A 1995 {\em \apj\/} {\bf 442}
  758--767

\bibitem[{{Wickramasinghe} \& {Ferrario}(2005)}]{WickramasingheFerrario05MN}
{Wickramasinghe} D~T and {Ferrario} L 2005 {\em \mnras\/} {\bf 356} 1576--1582

\bibitem[{{Auri{\`e}re} {et~al.}(2007){Auri{\`e}re}, {Wade}, {Silvester},
  {Ligni{\`e}res}, {Bagnulo}, {Bale}, {Dintrans}, {Donati}, {Folsom},
  {Gruberbauer}, {Bon Hoa}, {Jeffers}, {Johnson}, {Landstreet}, {L{\`e}bre},
  {Lueftinger}, {Marsden}, {Mouillet}, {Naseri}, {Paletou}, {Petit}, {Power},
  {Rincon}, {Strasser}, \& {Toqu{\'e}}}]{Auriereetal07}
{Auri{\`e}re} M, {\em et~al.~} 2007 {\em \aap\/}
  {\bf 475} 1053--1065 (\textit{Preprint} \eprint{0710.1554})

\bibitem[{{Shorlin} {et~al.}(2002){Shorlin}, {Wade}, {Donati}, {Landstreet},
  {Petit}, {Sigut}, \& {Strasser}}]{Shorlinetal02}
{Shorlin} S~L~S, {Wade} G~A, {Donati} J~F, {Landstreet} J~D, {Petit} P, {Sigut}
  T~A~A and {Strasser} S 2002 {\em \aap\/} {\bf 392} 637--652

\bibitem[{{Bagnulo} {et~al.}(2006){Bagnulo}, {Landstreet}, {Mason}, {Andretta},
  {Silaj}, \& {Wade}}]{Bagnuloetal06}
{Bagnulo} S, {Landstreet} J~D, {Mason} E, {Andretta} V, {Silaj} J and {Wade}
  G~A 2006 {\em \aap\/} {\bf 450} 777--791 (\textit{Preprint}
  \eprint{arXiv:astro-ph/0601516})

\end{thebibliography}

\end{document}